\documentstyle[12pt]{article}
\begin{document}
\baselineskip 16pt plus 2pt minus 2pt
\newcommand{\beq}{\begin{equation}}
\newcommand{\eeq}{\end{equation}}
\newcommand{\beqa}{\begin{eqnarray}}
\newcommand{\eeqa}{\end{eqnarray}}
\newcommand{\dfrac}{\displaystyle \frac}
\renewcommand{\thefootnote}{\#\arabic{footnote}}
\newcommand{\ve}{\varepsilon}
\newcommand{\krig}[1]{\stackrel{\circ}{#1}}
\newcommand{\barr}[1]{\not\mathrel #1}
\begin{titlepage}

%\hfill hep--ph/9411xxx

%{\bf{DRAFT, \today}}

%{\large {\bf CONFIDENTIAL}}

\begin{center}

\vspace{2.0cm}

{\large  \bf { NOVEL PION ELECTROPRODUCTION \\

LOW--ENERGY THEOREMS}}

\vspace{1.2cm}

{\large  V. Bernard$^{\ddag,1)}$, N. Kaiser$^{\diamond,2)}$,
 Ulf-G. Mei\ss ner$^{\dag,3)}$}

\vspace{0.7cm}

$^{\ddag}$Centre de Recherches Nucl{\'e}aires et Universit{\'e} Louis
Pasteur de  Strasbourg,\\ Physique Th{\'e}orique, BP 20Cr, F--67037
Strasbourg Cedex 2, France

$^\diamond$Technische Universit{\"a}t M\"unchen, Physik Department T30,
James--Franck--Stra{\ss}e, D--85747 Garching, Germany

$^{\dag}$Universit\"at Bonn, Institut f\"ur Theoretische Kernphysik, Nussallee
14-16,\\ D--53115 Bonn, Germany

\vspace{0.5cm}

email: $^{1)}$bernard@crnhp4.in2p3.fr, $^{2)}$nkaiser@physik.tu-muenchen.de,
$^{3)}$meissner@pythia.itkp.uni-bonn.de

\end{center}

\vspace{2.5cm}

\begin{center}

 ABSTRACT

\end{center}

\vspace{0.1cm}

\noindent
We present novel low--energy theorems for the P--wave multipoles
$2M_{1+} + M_{1-}$, $M_{1+} -M_{1-}$, $E_{1+}$ and $L_{1\pm}$ for neutral
pion electroproduction off protons. These should be very useful for the
analysis of existing or future threshold data.

\vspace{3.5cm}

\vfill

\noindent CRN 94--66 \hfill

\noindent TK 94 19 \hfill December 1994

\end{titlepage}

Over the last few years, very precise data
probing the structure of the nucleon
in  neutral pion photo-- and electroproduction off protons
in the threshold region  \cite{beck} \cite{pat} have become available.
At present, further data from MAMI at Mainz and NIKHEF at Amsterdam
for $\gamma^\star p \to \pi^0 p$ (here, $\gamma^\star$ denotes the virtual
photon)  are in the process of being analyzed \cite{walch} \cite{blok}.
These are characterized by a small photon four--momentum, $k^2 \simeq -0.1$
GeV$^2$, and $\Delta W = W - W_{\rm thr}$ of a few MeV, with $W$ the cms
energy of the pion--nucleon system in the final state. Even if one restricts
oneself to the approximation of retaining only the
S-- and P--wave multipoles, there are already seven of them.
Another complication arises from the fact that
certain of these multipoles are much bigger than some others.
Therefore, to get a good determination of all multipoles one does not only need
a set of very precise data but also model--independent constraints to perform a
reliable multipole analysis. We recall  that the determination of the
electric dipole amplitude $E_{0+}$ in neutral pion photoproduction off protons
based on the Mainz data \cite{beck} is very sensitive to the treatment of the
P--wave multipoles \cite{bh} \cite{berg} \cite{dt}
and has only recently been put on a firmer theoretical basis
\cite{vero} \cite{bkmpi0}. In this note, we wish to
 extend this analysis to $\pi^0$ electroproduction from protons
and show that there exists a set of very useful low--energy
theorems for certain P--waves which should be used in the data analysis. To
derive these, we explore the strictures of the chiral symmetry of QCD as
implemented in an effective field theory, here heavy baryon chiral perturbation
theory (HBCHPT) \cite{JM} \cite{BKKM}. We note that some of these results
were implicetely contained in the calculations presented in ref.\cite{bklm}
making use of relativistic baryon CHPT but have not been made explicit due to
the much more complicated arrangement of the chiral expansion in that framework
\cite{GSS}. For a general review, we refer to ref.\cite{UGM}.

To be specific, consider the process
$\gamma^\star (k) + p(p_1) \to \pi^0(q) + p(p_2)$ with
$k^2 < 0$ and $s = W^2 = (k+p_1)^2 = (q+p_2)^2$ the cms energy squared.
In the threshold region, i.e. when the pion three--momentum $\vec q$ is close
to zero, the transition matrix element $T \cdot \epsilon$ can be written in
terms of S-- and P--wave multipoles as follows:
$$ {m\over 4\pi \sqrt{s}} T\cdot \epsilon = \,  \,i \vec \sigma \cdot
\vec \epsilon \, [ E_{0+} + \hat q \cdot \hat k \, ( 3 E_{1+} + M_{1+}-
M_{1-})] + i \vec \sigma \cdot \hat k \,\vec \epsilon \cdot  \hat q \,( 3
E_{1+} - M_{1+}+ M_{1-})$$
$$ \quad + (\hat q \times \hat k) \cdot \vec \epsilon
\,(2 M_{1+}+ M_{1-}) + i \vec \sigma \cdot \hat k \, \vec \epsilon \cdot \hat
k \, [ L_{0+} - E_{0+} + 6\,\hat q \cdot \hat k \, ( L_{1+} - E_{1+})] $$

\beq \qquad \qquad
+ i \vec \sigma \cdot \hat q \,\vec \epsilon \cdot  \hat k \,(  L_{1-} -
2  L_{1+}) \eeq
in the $\pi N$ cm system and in the gauge $\epsilon_0 = 0$.
Gauge invariance $k \cdot T =0$ allows to recover $T_0$ via $T_0 = \vec k
\cdot \vec T \, / k_0$.
There are two S--wave multipoles, $E_{0+}$ and $L_{0+}$, and five P--wave
multipoles, $M_{1\pm}$, $E_{1+}$ and $L_{1\pm}$. Here $E,M$ and $L$ stand for
electric, magnetic and longitudinal, respectively, the first subscript gives
the $\pi N$ angular momentum and the $\pm$ refers to the
total angular momentum,
$j = l \pm 1/2$. The longitudinal ones,
$L_{0+} , L_{1\pm}$ are, of course, specific to electroproduction.  All
multipoles depend on $\omega$, the pion energy in the cm system, and on $k^2$.
We will suppress these arguments in what follows. Our aim is to give the
threshold expansion for the P--wave multipoles in powers of the dimensionless
parameters
\beq  \mu = \frac{M_\pi}{m}\, , \quad \rho = -\frac{k^2}{M_\pi^2} \,
\, , \eeq
using HBCHPT. Here, $M_\pi = 134.97$ MeV and $m = 928.27$ MeV are the
neutral pion and the proton mass, in order.
We will work within the one--loop approximation to order ${\cal
O}(p^3)$, where $p$ denotes a genuine small momentum. The corresponding chiral
counting rules are detailed e.g. in refs.\cite{ecker} \cite{weinnn}.
The effective Lagrangian
takes the form
\beq
{\cal L}_{\pi N} = {\cal L}_{\pi N}^{(1)} + {\cal L}_{\pi N}^{(2)} +
{\cal L}_{\pi N}^{(3)}   \, \, , \label{lpin} \eeq
where the superscript '(i)' ($i=1,2,3$) refers to the chiral dimension.
One--loop diagrams start at order $p^3$. The explicit form of the various terms
contributing to the process considered here can be found in ref.\cite{bkmpi0}
together with a determination of the corresponding low--energy constants which
appear at orders $p^2$ and $p^3$. In fact, only the anomalous magnetic moment
of the proton, $\kappa_p = 1.793$, and the constant $b_P = 15.8$ GeV$^{-3}$
will enter our results. The numerical value of $b_P$ can in fact be completely
understood by resonance exchange, with a share of 80$\%$ from the
$\Delta(1232)$ and 20$\%$ from vector meson exchange.

Let us first discuss the magnetic multipoles $M_{1+}$ and $M_{1-}$.  From
these, one forms the combinations $2M_{1+} + M_{1-}$ and $M_{1+} - M_{1-}$.
While the former is completely dominated by the contact term proportional to
$b_P$, the latter shows a rapidly converging expansion in $\mu$ \cite{conv}.
Straightforward calculation gives for the slopes of these particular
combinations of the magnetic multipoles at threshold:
\beq{1\over |\vec q \, | }\, ( 2M_{1+} +  M_{1-})^{\rm thr} =
 e M_\pi\sqrt{1+\rho}
\biggl( b_P + {g_{\pi N} \over 16 \pi m^3} \biggr) + {\cal O}(\mu^2)
\label{msum}  \eeq

$$ {1\over |\vec q \, | }\, ( M_{1+} - M_{1-})^{\rm thr} = {e g_{\pi N}
\over 8 \pi
m^2}  \sqrt{1+\rho} \biggl\{1 + \kappa_p + {\mu \over 4} \biggl[ - 5 - 2
\kappa_p  $$
\beq \qquad
+ {g_{\pi N}^2 \over 8 \pi} \biggl( {4+3\rho \over 1 +\rho} -
{(2+\rho)^2 \over 2 (1+\rho)^{3/2}} \arccos{-\rho \over 2+\rho} \biggl) \biggl]
\biggl\} + {\cal O}(\mu^2) \label{mdif} \eeq
with $g_{\pi N} = 13.4$ the strong pion--nucleon coupling
constant and $e^2 / 4 \pi = 1 / 137.036$.
The momentum dependence of eq.(\ref{msum}) is entirely given
by the square-root $\sqrt{1+\rho}$ arising from $|\vec k \,|$
whereas in eq.(\ref{mdif}) in addition
the function $4 - \Xi_1 (\rho)$ enters (cf. eq.(5.2)
in ref.\cite{bklm}). This is shown in fig.1a. We stress that this
momentum--dependence is not due to the proton magnetic form factor. Its
chiral expansion to order $p^3$ reads
\beq G_M^p(k^2) = 1 + \kappa_p + {g_{\pi N}^2 \over 32 \pi} \mu \biggl( 2 -
{4+\rho \over \sqrt{\rho}} \arctan{\sqrt{\rho}\over2}\biggr) + {\cal
O}(\mu^2) \, ,  \eeq
i.e. to the order we are working one is not sensitive to the on--shell
electromagnetic form factors. The $\rho$--dependent function
$4-\Xi_1(\rho)$ in eq.(\ref{mdif}) can  be considered as the leading
term in the $\mu$--expansion of the structure function related to the
half off--shell photon nucleon vertex (see the discusion in ref.\cite{bkm2}).
In the exact chiral limit $\mu = 0$ the expression in the curly bracket of
eq.(\ref{mdif}) becomes, however, equal to the proton magnetic
form factor (in the
chiral limit), $ 1 + \krig \kappa_p - (g^2_{\pi N}/64 \krig{m})
\sqrt{-k^2}$. Such a
behaviour is exactly required by the soft-pion theory.
The same features (in the chiral limit) were observed in
ref.\cite{bkms}
for double pion electroproduction at threshold.

For the slopes of the small multipoles $E_{1+}$ and $L_{1\pm}$ at
threshold, we find
\beq{1\over |\vec  q \, | }\, E_{1+}^{\rm thr} =
{e g_{\pi N}\mu \over 96 \pi m^2}
\sqrt{1+\rho} \biggl[1+ {g_{\pi N}^2 \over 8 \pi} \biggl( {8+5\rho \over 3(1
+\rho)^2} - {(2+\rho)^2 \over 2 (1+\rho)^{5/2}} \arccos{-\rho \over 2+\rho}
\biggl) \biggl] + {\cal O}(\mu^2) \label{e1} \eeq

\beq {1\over |\vec  q \, | }\, L_{1+}^{\rm thr} =
{e g_{\pi N}\mu \over 96 \pi m^2}
\sqrt{1+\rho} \biggl[1+ {g_{\pi N}^2 \over 16 \pi} \biggl( -{4+\rho \over (1
+\rho)^2} + {4-\rho^2 \over 2 (1+\rho)^{5/2}} \arccos{-\rho \over 2+\rho}
\biggl) \biggl] + {\cal O}(\mu^2)  \label{l1} \eeq

\beq {1\over |\vec  q \,| }\, L_{1-}^{\rm thr} =
{e g_{\pi N}\mu \over 12 \pi m^2}
\sqrt{1+\rho} \biggl[1+ {g_{\pi N}^2 \over 32 \pi} \biggl( {2 \over 1+\rho} -
{2+\rho \over  (1+\rho)^{3/2}} \arccos{-\rho \over 2+\rho}
\biggl) \biggl] + {\cal O}(\mu^2) \, \, . \label{l1m} \eeq
These are shown in fig.1b for $0 \le \rho \le 10$. We notice the rather
weak momentum dependence of $E_{1+}$ and of $L_{1+}$, i.e. there is
some balance between the square root (which multiplies the Born terms)
and the loop contribution  as given by the functions in the round
brackets of eqs.(\ref{e1},\ref{l1}). In case of the $L_{1-}$ multipole,
this loop--induced momentum dependence overtakes the $\sqrt{1+\rho}$
behaviour. Notice also that in $E_{1+}$ and $L_{1\pm}$ the negative
loop term is larger in magnitude than the small positive
Born contribution entering at the same chiral order.

Eqs.(\ref{msum},\ref{mdif},\ref{e1},\ref{l1},\ref{l1m}) constitute a
set of useful low--energy theorems \cite{gerulf} for neutral pion
electroproduction off protons and they should be used as constraints in
the analysis of the new and upcoming threshold data. The main features
of these results are, of course, not totally unexpected. The two
magnetic multipoles are dominant, and their $k^2$--dependence is
essentially governed by the  $\sqrt{1+\rho}$--factor with an extra
weak $\rho$--dependence for $M_{1+} - M_{1-}$ from the one--loop
graphs. The $E_{1+}$ and $L_{1\pm}$ multipoles are small corrections
to the dominant $M_{1\pm}$, their momentum--dependence shows a
stronger influence of the chiral loops. This latter effect is
particularly visible in $L_{1-}$, cf. fig.1b.

We have not yet made any statement about the two S--wave multipoles
$E_{0+}$ and $L_{0+}$. As detailed in refs.\cite{vero} \cite{bkmpi0}
\cite{bklm}, a higher order calculation is mandatory
since the corresponding series in $\mu$ are slowly converging.
At present, a precise experimental determination of
these S--wave multipoles is called for making use of the low--energy
theorems presented here. However, we remark here that the cusp effect
in $L_{0+}$ is expected to be weakened as $\rho$ increases in
comparison to the cusp in the electric dipole amplitude $E_{0+}$. To
be precise, to this order the corresponding imaginary parts read
\beq {\rm Im} E_{0+}(\omega,k^2) = {e g_{\pi N}  \over 32 \pi^2 F_\pi^2}
\mu \sqrt{\omega^2- \omega_c^2} + \dots  \eeq
\beq {\rm Im} L_{0+}(\omega,k^2) = {e g_{\pi N} \over 32 \pi^2
  F_\pi^2} \mu{M_\pi^2
\over 2M_\pi^2 - k^2 }  \sqrt{\omega^2-\omega_c^2} + \dots \, \, , \eeq
with $F_\pi = 93$ MeV the pion decay constant and $\omega_c =140.11$
MeV corresponding to the opening of the $\pi^+ n$ channel. The
strength of the imaginary part above $\omega_c$ is intimately
connected to the energy variation of the real part below $\omega_c$
(the cusp effect) as seen from dispersion theoretical considerations.
To leading order,  the cusp in  $E_{0+}$ is  indepedent of $k^2$ and
the corresponding cusp in $L_{0+}$ is suppressed by a factor of
$(2+\rho)^{-1}$, i.e. at $\rho = 3$ ($k^2 = -0.055$ GeV$^2$) it is diminished
to 20$\%$ of its magnitude in $E_{0+}$. Consequently,
one cannot expect to see such a cusp effect in $L_{0+}$ for typical
photon four--momenta of $|k^2| = 0.05 \ldots 0.10$ GeV$^2$. A further
crucial point in $\pi^0$ electroproduction is the $k^2$ dependence of
$E_{0+}^{\rm thr}$ and in particular the point where it changes
sign. We stress here that the low-energy theorem presented in
\cite{bklm} predicts a stronger increase (in $k^2$) than conventional
pseudovector Born calculations do (due to a particular large loop
effect at order $p^3$, i.e. the additional $\Xi_1 (\rho)$ term in
eq.(5.1) of ref.\cite{bklm}).
To be specific, the respective slope in $k^2$ reads
\beq {\partial E_{0+}^{\rm thr} \over \partial k^2}\bigg|_{ k^2 = 0}  = -
{e g_{\pi N}\over 16 \pi m^3} \biggl( 1 + \kappa_p + {4 - \pi \over 16 \pi}
{m^2 \over F_\pi^2} \biggr) + {\cal O}(\mu)  \eeq
where the last term in the bracket stems from the chiral loops. It amounts to
$62 \%$ correction to the (conventional) magnetic moment piece.
However, to draw decisive conclusions in
the S--waves, it seems necessary to perform higher order
calculations. We hope to come back to this topic in the near future.

\bigskip

\bigskip

\noindent {\Large{\bf Acknowledgements}}

\medskip

\noindent We thank Henk Blok for triggering our interest in this problem.

\bigskip

\bigskip

\bigskip

\noindent {\Large{\bf Figure Captions}}

\smallskip

\begin{enumerate}
\item[Fig.1a] The slope of the magnetic multipole combinations
$(2M_{1+}+M_{1-})/ |\vec q \,|$ (solid line) and
$(M_{1+}-M_{1-})/ |\vec q \,|$ (dashed line)
at threshold versus $\rho$ (in GeV$^{-2}$).
\item[Fig.1b] The slope of the small multipoles
$E_{1+}/ |\vec q \,|$ (solid line), $L_{1+}/ |\vec q \,|$ (dashed
line) and  $L_{1-}/ |\vec q \,|$ (dash--dotted line)
at threshold versus $\rho$
(in GeV$^{-2}$).
\end{enumerate}


\bigskip

\baselineskip 14pt

\begin{thebibliography}{99}

\bibitem{beck} R. Beck et al., Phys. Rev. Lett. {\bf 65} (1990) 1841

\bibitem{pat} T.P. Welch et al., Phys. Rev. Lett. {\bf 69} (1992) 2761

\bibitem{walch} Th. Walcher and M. Distler, private communication

\bibitem{blok} H. Blok, private communication

\bibitem{bh} A.M. Bernstein and B.R. Holstein,
Comments Nucl. Part. Phys. {\bf 20} (1991) 197

\bibitem{berg} J. Bergstrom, Phys. Rev. {\bf C44} (1991) 1768

\bibitem{dt} D. Drechsel and L. Tiator, J. Phys. G: Nucl. Part. Phys. {\bf 18}
(1992) 449

\bibitem{vero}
V. Bernard, ``Threshold Pion Photo-- and Electroproduction in Chiral
Perturbation Theory", talk given at the Workshop on Chiral Dynamics~:
Theory and
Experiment, MIT, Cambridge, July 1994, preprint CRN 94/45, hep-ph/9408323

\bibitem{bkmpi0}V. Bernard, N. Kaiser and Ulf-G. Mei\ss ner, ``Neutral
Pion Photoproduction off Nucleons Revisited'', preprint CRN 94-62
and TK 94 18, 1994, hep-ph/9411287

\bibitem{JM} E. Jenkins and A.V. Manohar, Phys. Lett. {\bf B255} (1991) 558

\bibitem{BKKM}
V. Bernard, N. Kaiser, J. Kambor and Ulf-G. Mei\ss ner, Nucl. Phys.
{\bf B388} (1992) 315

\bibitem{bklm} V. Bernard, N. Kaiser, T.--S. H. Lee
and Ulf-G. Mei{\ss}ner,  Phys. Rep. {\bf 246} (1994) 315

\bibitem{GSS} J. Gasser, M.E. Sainio and A. $\rm{\check S}$varc,
Nucl. Phys. {\bf B307} (1988) 779

\bibitem{UGM} Ulf-G. Mei{\ss}ner, Rep. Prog. Phys. {\bf 56} (1993) 903

\bibitem{ecker} G. Ecker, Czech. J. Phys. {\bf 44} (1994) 405

\bibitem{weinnn} S. Weinberg, Phys. Lett. {\bf B251} (1990) 288; Nucl. Phys.
{\bf 363} (1991) 3

\bibitem{conv} To be precise, this means that the term of order $\mu$ is much
smaller than the leading order one, $\sim 1+\kappa_p$,
 as long as $\rho$ does not become too large. For example, this correction
is 7$\%$ and 18$\%$ for $\rho = 1$ and $5$, respectively.

\bibitem{bkm2} V. Bernard, N. Kaiser and Ulf-G. Mei{\ss}ner,
Phys. Lett. {\bf B282} (1992) 448


\bibitem{gerulf} For a pedagocical discussion about the meaning of
low--energy theorems in the Standard Model, we refer the reader to
G. Ecker and  Ulf-G. Mei\ss ner, ``What is a
Low--Energy Theorem?'', preprint CRN 94--52 and UWThPh-1994-33, 1994,
to appear in Comments  Nucl. Part. Physics.

\bibitem{bkms} V. Bernard, N. Kaiser,  Ulf-G. Mei{\ss}ner and A. Schmidt,
Nucl. Phys. {\bf A580} (1994) 475



\end{thebibliography}
\end{document}